# Density-Dependent Electron Transport and Precise Modeling of GaN HEMTs

Sanyam Bajaj[1,a)], Omor F. Shoron[1], Pil Sung Park[1], Sriram Krishnamoorthy[1], Fatih Akyol[1], Ting-Hsiang Hung[1], Shahed Reza[2], Eduardo M. Chumbes[2], Jacob Khurgin[3] and Siddharth Rajan[1]

[1]Department of Electrical & Computer Engineering, The Ohio State University, Columbus, OH 43210

[2]Raytheon Integrated Defense Systems, Andover, MA 01810

[3]Department of Electrical & Computer Engineering, Johns Hopkins University, Baltimore, MD 21218

[4]Department of Material Science & Engineering, The Ohio State University, Columbus, OH 43210

**Abstract:** We report on the direct measurement of two-dimensional sheet charge density dependence of electron transport in AlGaN/GaN high electron mobility transistors. Pulsed IV measurements established increasing electron velocities with decreasing sheet charge densities, resulting in saturation velocity of $1.9 \times 10^7$ cm/s at a low sheet charge density of $7.8 \times 10^{11}$ cm$^{-2}$. A new optical phonon emission-based electron velocity model for GaN is also presented. It accommodates stimulated LO phonon emission which clamps the electron velocity with strong electron-phonon interaction and long LO phonon lifetime in GaN. A comparison with the measured density-dependent saturation velocity shows that it captures the dependence rather well. Finally, the experimental result is applied in TCAD-based device simulator to predict DC and small signal characteristics of a reported GaN HEMT. Good agreement between the simulated and reported experimental results validated the measurement presented in this report and established accurate modeling of GaN HEMTs.

---

[a)] Author to whom correspondence should be addressed. Electronic mail: bajaj.10@osu.edu Tel: +1-614-688-8458

III-nitride high electron mobility transistors (HEMTs) have demonstrated promising high frequency operation for RF power amplifying applications[1-4]. Their high current and power gain capability, and further increases in the operation frequency could enable high power electronics in the mm-wave and THz regime. The intrinsic current unity gain cutoff frequency ($f_{T,intrinsic}$) is directly proportional to the electron velocity in the channel. While parasitics do play an important role in the frequency performance, recent advances in ohmic contact technology have achieved ultra-low resistance[2,5,6] ($< 0.1$ $\Omega$.mm), thereby making the channel transit delay the main component of the overall delay in GaN HEMTs. Monte Carlo simulations in GaN have predicted peak electron drift velocities as high as 3 x $10^7$ cm/s, with comparatively lower saturation velocities up to ~ 2 x $10^7$ cm/s, limited mainly by optical phonon scattering[7-11]. From the total time delay analysis in recent ultra-scaled GaN HEMT results, the average electron velocity, $v_{ave}$, of the devices was found to be in the range from $1 \times 10^7$ to $2.8 \times 10^7$ cm/s[1-4].

Most reports of GaN HEMT RF performance report the peak cutoff frequency, which is typically observed at 10-20% of the highest saturated current density. The current gain, however, is found to drop significantly as the current density is increased. The transconductance profile in GaN HEMTs also shows a similar trend. For applications of GaN HEMTs in power amplifiers, where large signal gain and linearity are critical, the current and power gain values over the entire operating range are important. The reason for this reduction in current gain and transconductance as a function of channel charge density was investigated theoretically and shown to be directly related to the charge-dependent channel velocity[12]. In this report, we directly measure two-dimensional sheet density ($n_s$)-dependent electron velocity in AlGaN/GaN HEMTs at room temperature. A new carrier transport model based on the strong interaction between optical phonons and electron carriers in GaN HEMTs is also presented. Finally, the measured sheet density dependence of electron transport in two-dimensional electron gases (2DEGs) is applied in TCAD simulation of DC and RF characteristics of a reported GaN HEMT and compared to the experimental results.

HEMT structures (20 nm $Al_{0.23}Ga_{0.77}N$/GaN) shown in Figure 1, grown on SiC substrates by plasma-assisted molecular beam epitaxy (PAMBE)[13] were used for this study. I-shaped test structure consisting of wide contact/access regions and a thin constriction defined by mesa isolation (of depth 100 nm) was used to investigate the velocity field characteristics (Figure 2). Due to the geometry of this device, the resistance from the ohmic/access regions is significantly lower (at least 14X) than that from the constriction[14,15]. Therefore it can be assumed that entire potential drop and current limiting mechanisms are related to the constriction. The velocity $v_e$ was estimated from the current density J ($v_e = \frac{J}{qn_S}$) as a function of sheet charge density ($n_s$), where q is the electronic charge. The sheet charge density was estimated by Hall measurements made on van der Pauw patterns. Since the resistance of the constriction is estimated to be significantly higher than the ohmic/access regions, the field F across a constriction of length d was assumed to be given by $F = \frac{V}{d}$, where V is the applied bias. To measure velocity at varying electron concentrations, the constriction region was recess etched to different depths while the access regions were left unetched to maintain a low access resistance. All measurements were done at room temperature.

The fabrication of the test structures was done as follows. Ti/Al/Ni/Au ohmic contacts (100 μm wide) were deposited by e-beam evaporation and alloyed through rapid thermal annealing for 30 seconds at 850°C in $N_2$ ambient. $Cl_2$/$BCl_3$ plasma-based dry etching (RIE power of ~ 10 Watts) was used to define isolation mesas. For velocity measurements, I-shaped test structures of width 2 μm, 5 μm and 10 μm, and length 0.7 μm, 1 μm and 2 μm were used. The constriction dimensions were inspected using scanning electron microscope to confirm an error of ~ 2% as shown in Figures 2(c) and 2(d). To obtain different sheet charge densities, low-damage ICP-RIE etching was used (RIE power of ~ 3 Watts). Ohmic contact resistance of 0.4 Ω.mm was extracted using transmission length method measurements. For each recess depth, on-wafer Hall measurements were used to determine the sheet charge density and Hall mobility. The extracted values for electron sheet density and the corresponding mobility across several dies with different recess depths are shown in Figure 3. To measure the velocity-field curves, pulsed IV measurements were done using 2-terminal GS RF probes[16] and a commercial Dynamic IV Analyzer system[17]. A pulse duration of 500ns with a duty cycle of 0.01% was used. The conditions were chosen to ensure that self-heating effects did not have significant impact on the measured data. To eliminate leakage currents from the buffer layers and the instrument, an adjacent mesa-isolated device was measured and subtracted from each active device I-V characteristic.

Figure 4 shows an example of a series of pulsed measurements carried out at different sheet charge densities. The velocity profiles start to saturate above ~ 200 kV/cm, though the shape of the curves is different from different carrier density. There is negligible increase in the velocity beyond 500 kV/cm in all cases. At lower sheet charge density, the saturated velocity is significantly higher than that at higher sheet charge density. An increase in the noise level is also noticed with decreasing $n_s$ since the noise from the equipment is more evident as the device current density reduces. Using the velocity value at the maximum field (600 kV/cm) as the saturated velocity, the saturation velocity is plotted as a function of the 2DEG charge density. For each charge density, multiple devices were measured to verify uniformity, and the spread in the velocity is indicated through error bars. The saturation velocity at low sheet charge density is $1.9 \times 10^7$ cm/s, which is similar to that estimated by previous work in GaN[8,9,18]. However, as the sheet charge density is increased above $10^{13}$ cm$^{-2}$, the electron saturation velocity decreases significantly (by approximately 50%). As discussed later in this work, this velocity decrease plays a critical role in device characteristics, including the current density, transconductance profile and small-signal parameters.

Such a strong dependence of saturation velocity on the sheet charge density measured here cannot be explained by conventional mechanisms, such as band non-parabolicity and inter-valley transfer. One plausible cause may be related to hot phonons, or, more specifically to their stimulated emission. Previous works[19,20] have shown that the presence of hot phonons increases momentum relaxation time while not affecting the energy relaxation time – a clear recipe for the velocity saturation. The hot LO phonons are prevalent in wide gap nitride semiconductors (unlike Si and GaAs) due to their relatively long (few picoseconds) lifetimes[21] $\tau_{LO}$. These early models have shown qualitative agreement with a few experimental data points available at a time, but did not faithfully reproduce the clear "clamping" that can be observed in Figure 4. The reason for it was that references 19 and 20 used standard Fermi-Dirac distribution function for which, as shown in Figure 6(a), the electron population in the state with larger momentum $f_2$ is always less than population of the state with a smaller momentum $f_1$. Therefore, the probability of absorption of LO phonon with momentum $q = k_2 - k_1$ is always stronger than stimulated emission.

If, however one uses a more suitable model, with the so-called drifted Fermi-Dirac distribution[22] in which Fermi function is shifted by the amount of "average" drift momentum $k_d = m_c v_d / \hbar$, then for small drift velocity $v_d$ the situation would not change significantly from Figure 6(a), but as $v_d$ increases, the situation when $f_2 > f_1$ shown Figure 6(b) arises. Now the higher energy state $k_2$ has larger population than the lower one $k_1$, and, with probability of stimulated LO emission exceeding that of LO absorption, phonons now experience gain $g_{LO}$ (per unit of time) that is naturally proportional to the carrier density.

This situation is entirely analogous to the photon gain in laser medium, and, just as in a laser, threshold condition $g_{LO}\tau_{LO} > 1$ is reached when the phonon gain exceeds phonon decay. At this point the density of LO phonons starts growing exponentially and velocity saturates as all the additional power gets immediately transferred to LO generation, just as the upper level population in the laser gets clamped at threshold. Clearly, the higher is the carrier density, the earlier is the threshold reached and velocity saturates at a lower value, although the exact dependence of saturation velocity on sheet density is complicated as electron temperature also changes with increase in current density. Performing self-consistent simulations with drifted Fermi-Dirac distribution we have fitted the results into expression shown in Eq. (1) where $n_{s,0}$ is $1.8 \times 10^{13}$ cm$^{-2}$.

$$v_{Sat}(n_S) = \frac{10^7}{0.38 + \left(\frac{n_S}{n_{S,0}}\right)^{0.45}} \tag{1}$$

The theoretical curve is displayed in Figure 5 and shows good agreement with the measured results, suggesting that the model explains strong dependence of saturation velocity on carrier concentration, and captures the physics of 2DEG transport in GaN rather well.

To further verify the presented measurement results and model, the experimental sheet density-dependent saturation velocity characteristics were used in two-dimensional (2-D) TCAD simulator Silvaco ATLAS to simulate DC and RF characteristics of a reported GaN HEMT for comparison[23]. The simulated device was a standard sub-100 nm AlGaN/GaN HEMT structure on SiC substrate, with 20 nm of AlGaN barrier layer, 1nm of AlN interlayer and GaN buffer. The source-to-drain distance, $L_{SD}$, and the gate length, $L_G$, were 1.1 µm, and 60 nm, respectively, and the heights of gate, $h_{gate}$, was set to be 560 nm. The gate resistance, $R_g$, was 0.15 Ω•mm, and contact resistance, $R_c$, was 0.23 Ω•mm, as described in [23]. Figure 7 shows the comparison between the simulated and the experimental transfer characteristics, current gain profile, and the output characteristics. Excellent agreement between the simulated and the experimental results validates the measured density-dependent electron transport in 2DEGs of GaN HEMTs, and illustrates the accurate prediction of their DC and RF characteristics using the measurement. This shows that using a density dependent velocity model is critical for predictive models of AlGaN/GaN HEMTs.

In conclusion, we have measured the sheet density dependence of electron velocity in 2DEG of AlGaN/GaN HEMTs. The measured saturated velocity was found to have strong dependence on carrier density, and it reduced significantly with increasing density. A new optical phonon emission-based electron velocity model for GaN is presented and shown to capture the measured dependence rather well. The presented model accommodates stimulated LO phonon emission which clamps the electron velocity with strong electron-phonon interaction and long LO phonon lifetime in GaN. This clamping of the saturation velocity is found to be strongly dependent to electron carrier density in GaN channel. Finally, the measured density dependence of saturation velocity was applied in TCAD-based device simulator to simulate DC and small signal characteristics of a reported GaN HEMT with sub-100nm gate-length. Excellent agreement between the simulated and experimental results validated the measurement presented in this report and established accurate modeling of GaN HEMTs using the measured field and density dependence of electron velocity. The presented direct measurement of density-dependent electron velocity characteristics, new physics-based model, and precise simulation of DC and RF characteristics of GaN HEMTs have important applications in designing and enabling future III-Nitride transistors for high gain and linearity performance. The presented work could be used for minimizing the use of resources for the optimization of the transistor structure from the experimental studies, and can also be used to provide the close-fit benchmark for the upcoming high-frequency mm-Wave applications.

Acknowledgement:

The authors gratefully acknowledge the support from DATE MURI (ONR N00014-11-1-0721, Dr. Paul Maki), EXEDE MURI (ONR N00014-12-0976, Dr. Daniel Green and Dr. Paul Maki) and Raytheon Integrated Defense Systems, and the valuable help from Ye Shao and Prof. Wu Lu (The Ohio State University).

Figure captions:

**Figure 1:** AlGaN/GaN HEMT used for the study; (a) Structure schematic; and (b) the associated energy-band diagram.

**Figure 2:** The I-shaped test structure used for electron transport measurements using 2-terminal RF probe; (a) isometric view; (b) micrograph and schematic; (c) SEM images of 2 μm x 2 μm and (d) a 5 μm x 5 μm constriction structures.

**Figure 3:** Measured Hall mobility as a function of sheet charge density ($n_s$) for different recesses using low-power plasma etch. On-wafer Hall measurements were done on van der Pauw structures. Inset: measured values of sheet charge density and the associated values of Hall mobility and estimated AlGaN barrier thickness[24].

**Figure 4:** Measured sheet charge density ($n_s$) dependent electron velocity-field characteristics for AlGaN/GaN HEMT plotted for various $n_s$ values.

**Figure 5:** (a) Sheet charge density ($n_s$) dependent electron saturation velocity ($v_{sat}$) extracted from measured velocity-field characteristics. The error bars include the $v_{sat}$ values extracted from multiple devices. The data is compared to the presented optical phonon-based model.

**Figure 6:** (a) Fermi-Dirac distribution for low drift velocity $v_d$ showing that for two states $k_1$ and $k_2$ probability of phonon absorption always exceeds that of stimulated emission (b) Femi-Dirac distribution for large $v_d$ showing that LO phonons experience electron-density-dependent gain that will eventually cause the velocity saturation.

**Figure 7:** Comparison between the simulated and the experimental device characteristics from [23]; (a) transfer characteristics; (b) current gain profile; and (c) output characteristics.

1.

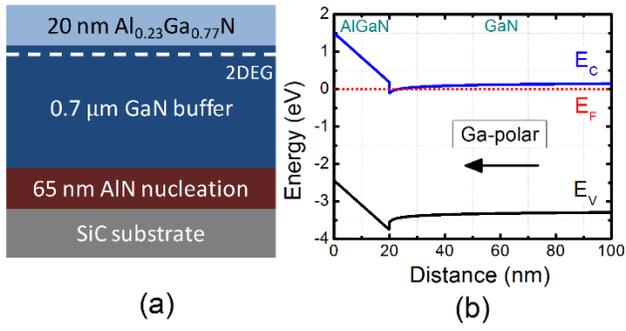

2.

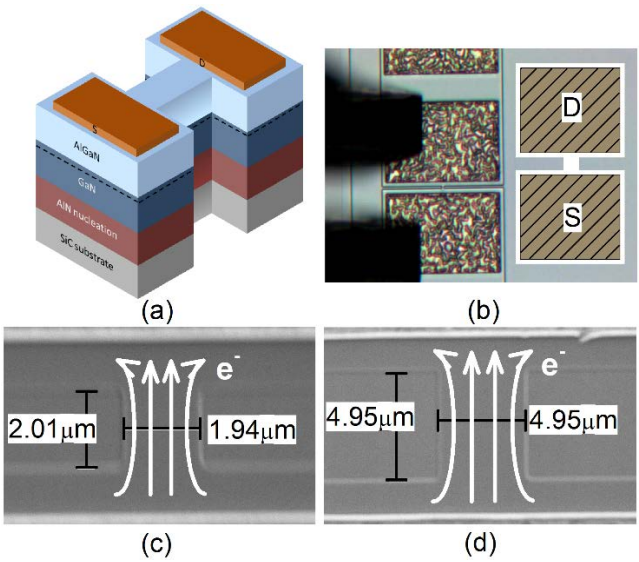

3.

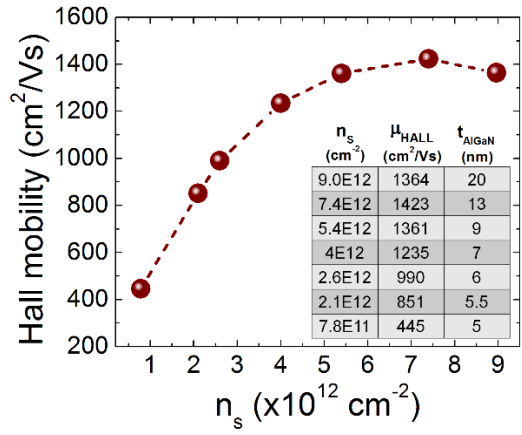

4.

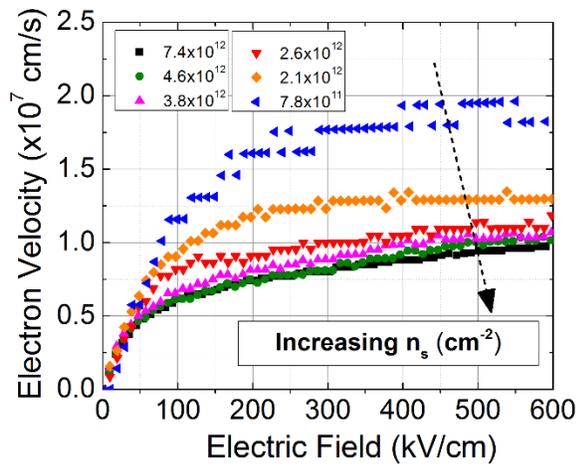

5.

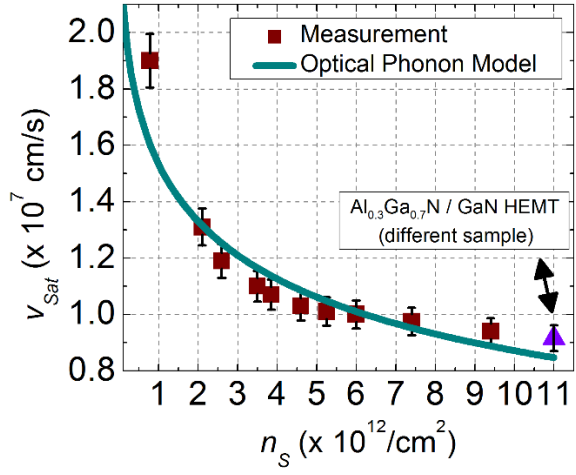

6.

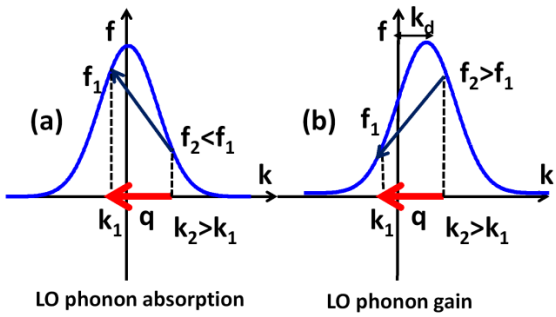

(a) LO phonon absorption

(b) LO phonon gain

7.

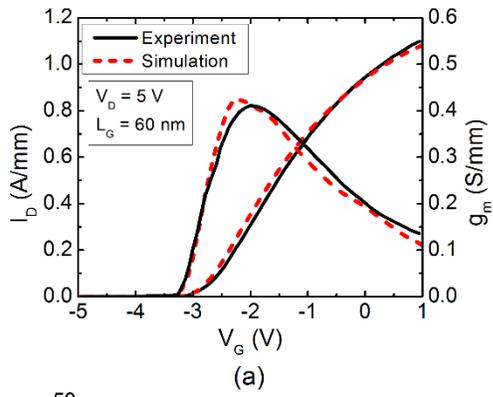

(a)

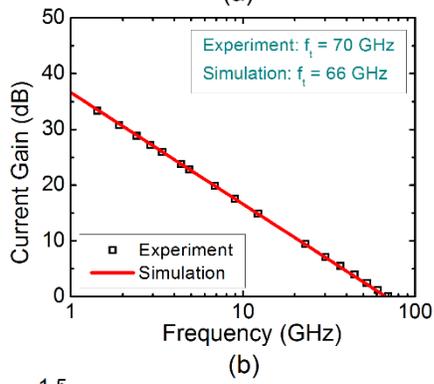

(b)

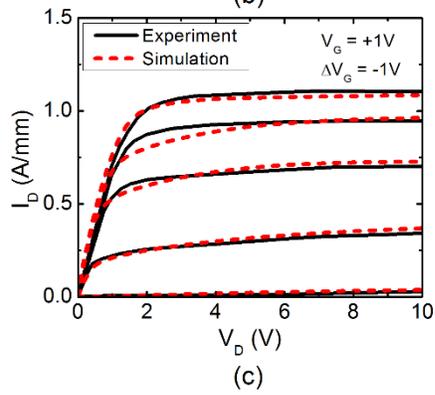

(c)